\documentstyle[12pt,a4,subeqn]{article}
\advance\voffset by 1.5truecm
\advance\hoffset by 1truecm
\begin{document}
\thispagestyle{empty}
\begin{flushright}
TIFR/TH/98-45 \\
November 1998 \\
\end{flushright}
\bigskip\bigskip\bigskip
\begin{center}
\Large{\bf MAXIMALLY CAUSAL QUANTUM MECHANICS}$^\star$ \\[1cm]
\large{S.M. ROY} \\[1cm]
{\sl Tata Institute of Fundamental Research,} \\ 
{\sl Homi Bhabha Road, Mumbai 400 005, INDIA} 
\end{center}
\vskip 5ex
\pretolerance=10000

\begin{abstract}

We present a new causal quantum mechanics in one and two dimensions
developed recently at TIFR by this author and V. Singh. In this theory
both position and momentum for a system point have Hamiltonian
evolution in such a way that the ensemble of system points leads to
position and momentum probability densities agreeing exactly with
ordinary quantum mechanics. 

\end{abstract}

\vspace{2in}


\vskip 3ex

\hrule width 3in

\medskip

\noindent $^\star$ Based on lecture presented at the Golden Jubilee
Workshop on Foundations of Quantum Theory, TIFR, 9--12 September,
1996.

\newpage

\noindent {\bf 1.}~ \underbar{\bf Nonexistence of History in Ordinary
Quantum Mechanics}.  At the 1927 Solvay conference Einstein discussed
the example of a particle passing through a narrow hole on to a
hemispherical fluorescent screen which records the arrival of the
particle.  Suppose that a scintillation is seen at a point $P$ at time
$t = T$, and suppose that the hole is so narrow that the wave packet
corresponding to the particle is uniformly spread all over the screen
at $t$ slightly less than $T$.  Was the particle somewhere near $P$ at
$t = T - \epsilon$ ($\epsilon$ small)?  Ordinary quantum mechanics
says that the probabilities at $t = T - \epsilon$ for the particle
being anywhere on the screen are uniform (and not particularly large
in the vicinity of $P$).  Thus the naive history corresponding to the
idea of a particle with a trajectory (\underbar{any} trajectory) is
denied. 

There have been recent attempts to define `consistent histories' in
quantum mechanics of open systems$^1$.  Apart from detailed features
found unattractive by some$^2$, there is the basic proposition by the
authors themselves that only very special sets of histories are
`consistent', and only these can be assigned probabilities.  For
example, in a double slit interference experiment we cannot assign a
probability that the particle reached a region of the screen having
earlier passed through slit 1 (except in the case of vanishing
interference). 
\bigskip

\noindent {\bf 2.} ~\underbar{\bf Lack of Causality in Ordinary
Quantum Mechanics}.  One of the definitions of causality (e.g. that
advocated by Jauch) is that ``Different results should have different
causes''.  Consider a quantum superposition $\alpha|+\rangle +
\beta|-\rangle$ for a spin-1/2 particle, where $|+\rangle$ and
$|-\rangle$ are eigenstates of $\sigma_z$ with eigenvalues +1 and -1
respectively.  When this state is prepared repeatedly and passed
through a Stern-Gerlach apparatus to measure $\sigma_z$, a fraction
$|\alpha|^2$ of the particles ends up at the detector corresponding to
$\sigma_z = +1$, and a fraction $|\beta|^2$ goes to the detector
corresponding to $\sigma_z = -1$.  Thus different results (going to
one detector or the other) arise from exactly the same cause (the same
initial state).  Of course this lack of causality might be restored in
a theory in which the wave function is not a complete description of
the state of the system.
\bigskip

\noindent {\bf 3.} ~\underbar{\bf Context Dependence of Quantum
Reality}.  In quantum mechanics, 
\[
|\psi(\vec x,t)|^2 d\vec x
\] 
is the probability of `observing' position to be in $d\vec x$ if
position were measured.  It is not the probability of position `being'
in $d\vec x$ independent of observation.  In fact, the same state
vector also yields
\[
|\tilde \psi (\vec p,t)|^2 d\vec p
\]
which is the probability of observing momentum to be in the interval
$d\vec p$ if momentum were to be measured.  The standard dogma is that
a simultaneous measurement of position and momentum is not possible.
For, if it were possible, it would collapse the state vector into a
simultaneous eigenstate of position and momentum which does not
exist.  Thus, quantum mechanics does not give probabilities for
position and momentum in the same experimental situation or
`\underbar{context}'.  Moreover, quantum mechanics cannot be embedded
in a stochastic hidden variable theory in which `reality' is context
independent$^3$.  Consider for example Bell's theorem$^3$ in the
context of Einstein-Podolsky-Rosen type measurements of $\vec \sigma_1
\cdot \vec a \ \vec \sigma_2 \cdot \vec b$ for a system of two spin
1/2 particles in the singlet state.  It shows that even if the
measurements of $\vec \sigma_1 \cdot \vec a$ and $\vec \sigma_2 \cdot
\vec b$ are made at spacelike separation, statistical predictions of
quantum mechanics are inconsistent with the assumption that the measured
value of $\vec \sigma_1 \cdot \vec a$ has a reality that is
independent of whether $\vec \sigma_2 \cdot \vec b$ or $\vec \sigma_2
\cdot {\vec b}'$ is measured together with it.  In this example
context dependence takes the form of violation of Einstein locality.
\bigskip

\noindent {\bf 4.} ~\underbar{\bf Causal Quantum Mechanics of de
Broglie and Bohm}.  De Broglie and Bohm$^4$ (dBB) proposed a theory
with position as a `hidden variable' so that $\{\vec
x,|\psi\rangle\}$, i.e., the state vector supplemented by the
instantaneous position is the complete description of the state of the
system.  Here $\vec x = (\vec x_1, \cdots, \vec x_N)$ denotes the
configuration space co-ordinate which evolves according to 
\begin{equation}
{d\vec x_i \over dt} = {1 \over m_i} \vec \nabla_i S(\vec x(t),t),
\end{equation}
where $m_i$ denotes the mass of particle $i$, and the Schr\"odinger
wave function is given by,
\begin{equation}
\langle \vec x|\psi(t)\rangle \equiv R \exp i S,
\end{equation}
with $R$ and $S$ real functions of $(\vec x,t)$.  DBB show that if we
start at $t=0$ with an ensemble of particles whose position density
coincides with $|\psi(\vec x,0)|^2$ at $t = 0$, and evolves with time
according to (1), then the position density coincides with $|\psi (\vec
x , t) |^2$ at any arbitrary time $t$. Thus, the phase space density
is
\begin{equation}
\rho_{dBB} (\vec x , \vec p , t) = |\psi (\vec x, t)|^2 \delta \left(\vec p
- \vec \nabla S (\vec x, t)\right)
\end{equation}
whose marginal at arbitrary time reproduces the position probability
density, 
\begin{equation}
\int \rho_{dBB} (\vec x, \vec p , t) \ d \vec p = | \psi (\vec x, t)
|^2 .
\end{equation}
Further, the time evolution (1) corresponds to evolution according to
a $c$-number causal Hamiltonian $H_c (\vec x, \vec p, t)$ in which the
potential has an added term ``the quantum potential'' which depends on
the wave function.

As far as the position variable is concerned the $dBB$ theory restores
history and causality without altering the statistical predictions of
quantum mechanics. The lack of Einstein locality is however an
essential feature of quantum mechanics. It is in sharp focus in
Eq. (1): the velocity of the $i$th particle depends on the
instantaneous position of all the particles however far they may be.

The momentum and other variables besides position do not have the same
favoured status as position however. As Takabayasi$^5$ pointed out the
$dBB$ phase space density does not yield the correct quantum momentum
density, i.e.,
\[
\int \rho_{dBB} (\vec x, \vec p, t) d \vec x \neq |\tilde\psi (\vec p,
t)|^2 .
\]
To overcome this problem $dBB$ introduce a measurement interaction
whose purpose is to convert the preexisting momentum prior to
measurement into one whose distribution agrees with the quantum
distribution. In contrast, for position, the value observed is the
same as the preexisting value. `Momentum' therefore has not the same
reality as `Position'.
\bigskip

\noindent {\bf 5.} ~\underbar{\bf Maximally Realitic Causal
Theory}. We asked the question$^{6,7}$, is it possible to remove this
asymmetrical treatment of position and momentum and build a new causal
quantum mechanics in which momentum and position can have simultaneous
reality? In Ref. 6, 7 we spelt out an affirmative answer in one
dimensional configuration space. here we recall the one dimensional
construction and also give the two dimensional generalization.

The point of departure is to seek a phase space density of the form
\begin{equation}
\rho (x, p, t) = |\psi (x, t)|^2 \delta (p - \hat p (x, t)) ,
\end{equation}
where $\hat p (x, t)$ is not given by the $dBB$ formula. Rather, $\hat
p (x, t)$ is to be determined by the requirement 
\begin{equation}
\int \rho (x, p, t) dx = | \tilde \psi (p, t) |^2 .
\end{equation}
If we assume that $\hat p (x, t)$ is a monotonic function of $x$
(non-decreasing or non-increasing),
\begin{equation}
\delta (p - \hat p (x, t)) = \frac{\delta (x - \hat x (p, t))} {\left|
\frac{\partial \hat p (x, t)} {\partial x}\right|}
\end{equation}
and
\begin{equation}
\rho (x, p, t) = \frac{|\psi (x, t)|^2}{\left| \frac{\partial \hat p
(x, t)} {\partial x} \right|} \delta (p - \hat p (x, t))
\end{equation}
If we determine $\hat p (x, t)$ such that
\begin{equation}
|\psi (x, t)|^2 = \left| \frac{\partial \hat p (x, t)}{\partial
x}\right| |\tilde\psi (p, t)|^2 ,
\end{equation}
we obtain
\begin{equation}
\rho (x, p, t) = | \tilde\psi (p, t) |^2 \ \delta (p - \hat p (x, t))
\end{equation}
which obeys the desired Eq. (6). Two explicit solutions to Eq. (9),
corresponding to non-decreasing $(\epsilon = 1)$ and non-increasing
$(\epsilon = -1)$ functions $\hat p (x, t)$ are given by,
\begin{equation}
\int^{\hat p (x, t)}_{-\infty} dp' |\tilde\psi (p', t)|^2 =
\int^{\epsilon x}_{-\infty} dx' |\psi (\epsilon x', t)|^2  .
\end{equation}
Instead of Eq. (5) or Eq. (10), the phase space density may now be
written in the symmetric form,
\begin{equation}
\rho (x, p, t) = |\psi (x,t)|^2 |\tilde\psi (p, t)|^2 \delta \left(
\int^p_{-\infty} dp' |\tilde\psi (p',t)|^2 - \int^{\epsilon
x}_{-\infty} dx' |\psi (\epsilon x', t)|^2 \right) .
\end{equation}
We have shown$^{6,7}$ that this phase space density corresponds to
evolution of position and momentum according to a $c$-number causal
Hamiltonian of the form
\begin{equation}
H_c (x,p,t) = \frac{1}{2m} (p - A (x,t))^2 + V (x,t) ,
\end{equation}
with both $A(x,t)$ and $V(x,t)$ having parts which depend on the wave
function. Thus we have two quantum potentials instead of just one
in the $dBB$ theory. The existence of $H_c$ ensures that the phase
space density obeys the Liouville condition. In one dimension it turns
out that
\[
\frac{dx}{dt} = \left(\frac{dx}{dt}\right)_{dBB},
\]
but
\[
\hat p (x,t) = m \frac{dx}{dt} - A (x,t) \neq \left( m
\frac{dx}{dt}\right)_{dBB} .
\]

It may be recalled that like the phase space distribution (12), the
Wigner distribution$^8$ also reproduces $|\psi (x,t)|^2$ and $|\tilde
\psi (p, t)|^2$ as marginals. However the Wigner distribution is not
positive definite and cannot therefore have a probability
interpretation; further the condition of Hamiltonian evolution of $x,
p$ and the consequent Liouville property are valid for the phase space
density (12) but not for the Wigner distribution. 

In higher dimensions there is a surprise. E.g. for $n=2$, we can
construct a causal quantum mechanics in which the quantum probability
densities corresponding to three different complete commuting sets (CCS) of
observables, e.g. $(X_1, X_2), (P_1, X_2), (P_1, P_2)$ is
simultaneously realized. Explicitly, the positive definite phase space
density
\begin{equation}
\rho (\vec x, \vec p, t) = |\psi (x_1, x_2, t)|^2 |\psi (p_1, x_2,
t)|^2 |\psi (p_1, p_2, t)|^2 \delta (A_1) \delta (A_2)
\end{equation}
where
\begin{eqnarray}
A_1 &\equiv& \int^{p_1}_{-\infty} |\psi (p'_1, x_2, t)|^2 dp'_1 -
\int^{x_1}_{-\infty} |\psi (x'_1, x_2, t)|^2 dx'_1 , \\
A_2 &\equiv& \int^{p_2}_{-\infty} |\psi (p_1, p'_2, t)|^2 dp'_2 -
\int^{x_2}_{-\infty} |\psi (p_1, x'_2, t)|^2 dx'_2 ,
\end{eqnarray}
reproduces as marginals the correct quantum probability densities
$|\psi (x_1, x_2, t)|^2 , |\psi (p_1, x_2, t)|^2$ and $|\psi (p_1,
p_2, t)|^2$. (For notational simplicity we have omitted the tildas
denoting Fourier transforms).  The corresponding velocities are
however in general 
different from the $dBB$ velocities. The calculation of the
velocities, and the explicit causal Hamiltonian for $n \geq 2$ will be
given in detail in a later publication$^9$ where we show that the
quantum probability densities corresponding to $n+1$ CCS can be
simultaneously realized (i.e., with one phase space density). A recent
quantum phase space contextuality theorem proves that$^{10}$ it is
impossible to simultaneously realize quantum probability densities
corresponding to all possible CCS. We conjecture that the causal
quantum mechanics we have constructed is maximally realistic. 

\newpage

\noindent{\bf References} \\

\begin{enumerate}
\item[{1.}] R.B. Griffiths, {\it J. Stat. Phys.} \underbar{36}, 219
(1984); {\it Phys. Rev. Lett.} \underbar{70}, 2201 (1993);
M. Gell-Mann and J.B. Hartle, in Proc. 25th Int. Conf. on High Energy
Physics, Singapore 1990, eds. K.K. Phua and Y. Yamaguchi (World
Scientific, 1991); R. Omn\'es, {\it Rev. Mod. Phys.} \underbar{64},
339 (1992) and ``\underbar{The Interpretation of Quantum Mechanics}''
(Princeton Univ. Press 1994).
\item[{2.}] F. Dowker and A. Kent, {\it J. Stat. Phys.}
\underbar{82}, 1575 (1996); {\it Phys. Rev. Lett.} \underbar{75}, 3038
(1995).
\item[{3.}] A.M. Gleason,  {\it J. Math. \& Mech.} \underbar{6}, 885
(1957); S. Kochen and E.P. Specker, {\it J. Math. \& Mech.}
\underbar{17}, 59 (1967); J.S. Bell, {\it Physics} \underbar{1}, 195
(1964); A. Martin and S.M. Roy, {\it Phys. Lett.} \underbar{B350},
66 (1995).
\item[{4.}] L. de Broglie, ``Nonlinear Wave Mechanics, A Causal
Interpretation'', (Elsevier 1960); D. Bohm, {\it Phys. Rev.}
\underbar{85}, 166; 180 (1952); D. Bohm and J.P. Vigier, {\it
Phys. Rev.} \underbar{96}, 208 (1954). 
\item[{5.}] T. Takabayasi, {\it Prog. Theor. Phys.} \underbar{8}, 143
(1952).
\item[{6.}] S.M. Roy and V. Singh, `Deterministic Quantum Mechanics in
One Dimension', p. 434, Proceedings of International Conference on
Non-accelerator Particle Physics, 2-9 January, 1994, Bangalore, Ed. R.
Cowsik (World Scientific, 1995).
\item[{7.}] S.M. Roy and V. Singh {\it
Mod. Phys. Lett.} \underbar{A10}, 709 (1995).
\item[{8.}] E. Wigner, {\it Phys. Rev.} \underbar{40}, 749 (1932).
\item[{9.}] S.M. Roy and V. Singh, preprint TIFR/TH/98-42.
\item[{10.}] A. Martin and S.M. Roy, {\it Phys. Lett.} \underbar{B350},
66 (1995).
\end{enumerate}

\end{document}